\documentclass{iaus}
\usepackage{graphicx}

\title[S262.~~Fitting the 6dF Fundamental Plane] 
{Maximum\,likelihood\,method\,for\,fitting\,the Fundamental\,Plane\,of\,the\,6dF\,Galaxy\,Survey}

\author[C. Magoulas, M. Colless, D. H. Jones, C. M. Springob \& J. R. Mould]   
{Christina Magoulas$^1$, Matthew Colless$^2$, D. Heath Jones$^2$, Christopher M.
Springob$^2$ \and Jeremy R. Mould$^1$}

\affiliation{$^1$School of Physics, University of Melbourne, Australia \\[\affilskip]
$^2$Anglo-Australian Observatory, Epping, Australia}

\pubyear{2009}
\volume{262}  
\pagerange{1--3}
\jname{Stellar Populations: Planning for the Next Decade}
\editors{S. Charlot, G. Bruzual, eds.}
\begin{document}

\maketitle

\begin{abstract}

We have used over 10,000 early-type galaxies from the 6dF Galaxy Survey (6dFGS)  to construct the Fundamental Plane across the optical and near-infrared passbands. We demonstrate that a maximum likelihood fit to a multivariate Gaussian model for the distribution of galaxies in size, surface brightness and velocity dispersion can properly account for selection effects, censoring and observational errors, leading to precise and unbiased parameters for the Fundamental Plane and its intrinsic scatter. This method allows an accurate and robust determination of the dependencies of the Fundamental Plane on variations in the stellar populations and environment of early-type galaxies. 

\end{abstract}

\firstsection 

\section{Introduction}

One of the most important empirical scaling relations of early-type galaxies is the Fundamental Plane - a  tight correlation in the log space of effective radius $R_{e}$, central velocity dispersion $\sigma$,  and mean effective surface brightness $\langle \mu_{e} \rangle$  (\cite[Dressler et al. 1987]{Dressler_etal87}). 

Fundamental Plane studies aim to understand what physical processes constrain early-type galaxies to this plane rather than being uniformly scattered.  However such studies raise as many questions as they answer, not least being how the Fundamental Plane varies with stellar population and environment, and the origin of the observed \emph{tilt} of the plane with respect to virial expectations.  

These questions will be addressed in our analysis of early-type galaxies from the 6dF Galaxy Survey.  As well as determining the Fundamental Plane across a range of masses and environments, we will in future determine precise Fundamental Plane distances and peculiar velocities with which to map the local peculiar velocity field out to $\sim$16,500 km\,s$^{-1}$.

Recent studies have found various correlations of the Fundamental Plane with age and stellar population (\cite[Graves et al. 1987]{Graves_etal09}, \cite[Gargiulo et al. 2009]{Gargiulo_etal09}).  Therefore, as part of the 6dFGS Fundamental Plane study we will measure Lick indices to give ages, metallicities and $\alpha$-enhancements and quantify the effect of stellar populations on the Fundamental Plane.  We will investigate age-metallically effects and observe any trends in the mass-to-light ratio along the Fundamental Plane.  
\clearpage
\section{The 6dF Galaxy Survey}

The 6dF Galaxy Survey (\cite[Jones et al. 2004]{Jones_etal04}) is a combined redshift and peculiar velocity survey that covers the whole southern sky with the exception of the region within 10$^{\circ}$ of the Galactic Plane (i.e. $|$b$| > $ 10$^{\circ}$). The redshift survey contains measurements of more than 150,000 redshifts with a median of 0.053 (\cite[Jones et al. 2009]{Jones_etal09}).

The subsample of the brightest galaxies with high signal-to-noise (required for accurate velocity dispersions) constitutes the peculiar velocity sample.  The 6dFGS galaxies were primarily near-infrared-selected from the Two Micron All-Sky Survey (2MASS) Extended Source Catalogue (XSC) (\cite[Jarrett et al. 2000]{Jarrett_etal00}).  Near-infrared wavelengths are less sensitive to dust extinction, allowing measurements closer to the Galactic Plane, and also are closer to a stellar-mass selected sample.  This has the advantage of favouring early-type galaxies (rather than the younger, star-forming galaxies) and is therefore ideal for Fundamental Plane studies.  The peculiar velocity sample is expected to contain $\sim$10,000 galaxies out to a redshift of $\sim$16,500 km\,s$^{-1}$. 6dFGS provides one of the largest and homogeneous Fundamental Plane samples due to its large volume and uniform coverage.

\section{Method and Results}

The accuracy with which the Fundamental Plane is determined, is highly dependent on how well the fitting method used accounts for the intrinsic scatter and distribution of the sample. Therefore the 6dFGS Fundamental Plane is fit in $\log(R_{e})-\log(\sigma)-\langle \mu_{e} \rangle$ space with a tri-variate Gaussian distribution, and best-fit parameters are determined using the method of maximum likelihood.  We model the galaxy sample as a tri-variate Gaussian, as empirically it is a better representation of the data than a uniformly populated plane with Gaussian scatter. 

This maximum likelihood fitting procedure also properly accounts for the observational errors (and any correlations) in all galaxy observables, as well as selection criteria in velocity dispersion, magnitude and others variables. Using this fitting method is crucial in obtaining an unbiased view of the Fundamental Plane, unlike previous studies that do not properly account for all of these effects.

\subsection{Environmental Dependence}

The 6dFGS galaxies cover the whole southern sky and a full range of environments and are found in a uniform group/cluster catalogue (\cite[Merson et al. 2009]{Merson_etal09}).  Using this catalogue we can therefore investigate any environment properties of the Fundamental Plane.  

\begin{figure}[t]
\begin{center}
 \includegraphics[width=1.0\textwidth]{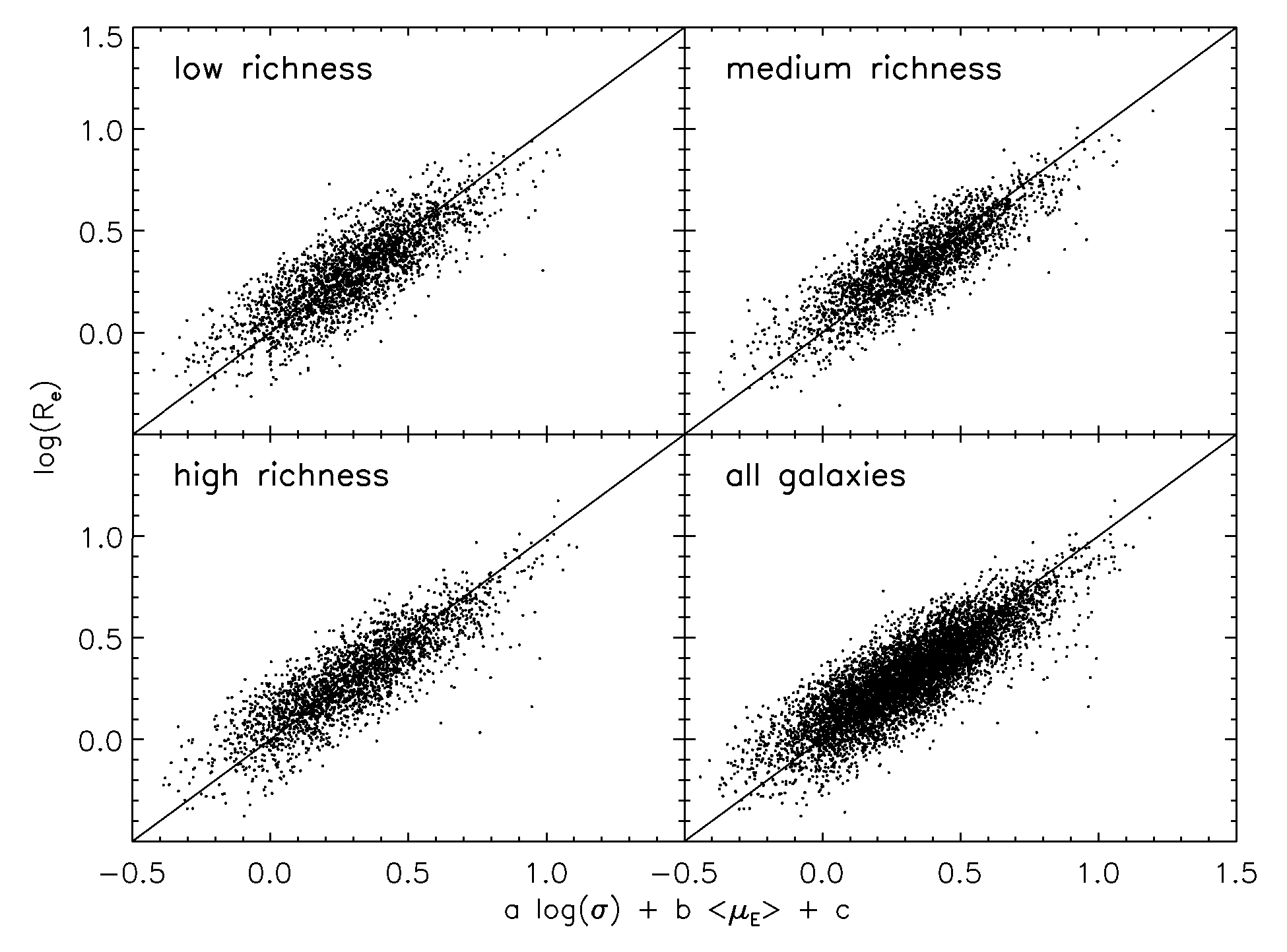} 
 \caption{The best-fit 6dFGS J-Band Fundamental Plane using maximum likelihood fitting for three richness subsamples (low richness - 3897 galaxies; medium richness - 3177 galaxies; high richness - 3126 galaxies) and the total sample consisting of 10200 galaxies.  In each case there is no significant variation of the Fundamental Plane in the different richness subsamples as all preliminary fits give coefficients of a$\sim$1.4,  b$\sim$0.30 and c$\sim$-8.15 and have similar dispersions about the plane of $\sigma\sim$0.06 } 
   \label{fig1}
\end{center}
\end{figure}

Initially we divide the sample into three richness subsamples defined as high richness (galaxies mostly in clusters), medium richness (galaxies mostly in groups) and low richness (singleton galaxies).  Our preliminary fit (Fig.\,\ref{fig1}) to these subsamples suggests there is no apparent variation of the Fundamental Plane with environment (\cite[Magoulas et al. 2009]{Magoulas_etal09}).  Although there may be variations with local density yet to be explored, this result has important ramifications for the 6dFGS peculiar velocity study as it becomes difficult to measure distances using groups and clusters if the Fundamental Plane in different environments has a different slope.  Hence, ensuring the Fundamental Plane is homogeneous across environments is paramount for its use in mapping the peculiar velocity field.

\clearpage

\end{document}